# The role of the dissipative and random forces in the calculation of the pressure of simple fluids with dissipative particle dynamics

A. Gama Goicochea[*], M. A. Balderas Altamirano, J. D. Hernández, and E. Pérez

Instituto de Física, Universidad Autónoma de San Luis Potosí, Álvaro Obregón 64, 78000, San Luis Potosí, SLP, Mexico

**ABSTRACT**

The role of viscous forces coupled with Brownian forces in momentum – conserving computer simulations is studied here in the context of their contribution to the total average pressure of a simple fluid as derived from the virial theorem, in comparison with the contribution of the conservative force to the total pressure. The specific mesoscopic model used is the one known as dissipative particle dynamics, although our conclusions apply to similar models that obey the fluctuation – dissipation theorem for short range interactions and have velocity – dependent viscous forces. We find that the average contribution of the random and dissipative forces to the pressure is negligible for long simulations, provided these forces are appropriately coupled and when the finite time step used in the integration of the equation of motion is not too small. Finally, we study the properties of the fluid when the random force is made equal to zero and find that the system freezes as a result of the competition of the dissipative and conservative forces.

---

[*]Corresponding author. Electronic mail: agama@alumni.stanford.edu

**INTRODUCTION**

Current research in soft matter systems requires of increasingly sophisticated computational tools to solve accurately models with large number of particles and complex interactions in the shortest possible time, so that the predictions can be used to interpret experimental trends, test effective theories and design new materials [1]. There are important reasons that have made of computer simulations the successful tools that they are today, among which are the fact that the interactions in many – body systems can be solved almost exactly, while most analytical theories must rely on approximations [2]. Additionally, one has total control over the thermodynamic and physicochemical conditions of the simulations, which is hardly achieved in most experiments. One of the most successful techniques used in the past decade or so for the modeling of complex fluids at the coarse – grained level is the so called dissipative particle dynamics (DPD) method [3, 4]. The model consists of three different types of forces that act between pairs of particles, namely a conservative force which is responsible for the thermodynamic properties of the fluid, and dissipative and random forces that are coupled in a way that keeps the temperature constant [4]. The forces are central and obey Newton´s third law, which leads to the local and global conservation of momentum. This aspect of DPD is very useful when modeling situations where hydrodynamic modes may play a leading role [5] because such modes are conserved, which is not necessarily the case when other types of dynamics with viscous and Brownian forces are used [6].

DPD has shown to be successful for the study of the properties of polymer blends [7], complex fluids under confined geometries [8], systems of biological [9] or industrial [10] interests, and for the prediction of properties of fluids under the influence of external stimuli such as steady flow [11], pH [12], or temperature [13]. In view of the numerous examples where DPD has been used to predict correctly the behavior of soft matter systems it is important to review some of the assumptions of the model. The purpose of this work is to determine quantitatively the extent to which the viscous and stochastic forces in DPD perturb the dynamics of the system through their contribution to the total pressure of the fluid via the virial theorem, in comparison with the contribution of the conservative force. This has not yet been tested explicitly and a quantitative understanding of this aspect is necessary before a fully satisfactory knowledge of DPD can be achieved.

**THE DPD MODEL**

The statistical mechanical basis of DPD was laid out in the work of Español and Warren [4]; following the original proponents of DPD [3] they modeled the dissipative force ($\boldsymbol{F}_{ij}^D$) and the random force ($\boldsymbol{F}_{ij}^R$) as

$$\boldsymbol{F}_{ij}^D = -\gamma \omega^D(r_{ij})[\hat{\boldsymbol{r}}_{ij} \cdot \boldsymbol{v}_{ij}]\hat{\boldsymbol{r}}_{ij} \qquad (1)$$

$$\boldsymbol{F}_{ij}^R = \sigma \omega^R(r_{ij})\xi_{ij}\hat{\boldsymbol{r}}_{ij} \qquad (2)$$

where $\boldsymbol{r}_{ij} = \boldsymbol{r}_i - \boldsymbol{r}_j$, $r_{ij} = |\boldsymbol{r}_{ij}|$, $\hat{\boldsymbol{r}}_{ij} = \boldsymbol{r}_{ij}/r_{ij}$; $\boldsymbol{r}_{ij}$ is the relative position vector between particles $i$ and $j$, $\sigma$ is the noise amplitude, $\gamma$ is the viscous force amplitude and $\boldsymbol{v}_{ij} = \boldsymbol{v}_i - \boldsymbol{v}_j$ is the relative velocity between the particles, with $\xi_{ij} = \xi_{ji}$ being random numbers with Gaussian distribution between 0 and 1, and unit variance. These numbers obey the following relations [14]:

$$\langle \xi_{ij}(t) \rangle = 0 \qquad (3)$$
$$\langle \xi_{ij}(t)\xi_{kl}(t') \rangle = \big(\delta_{ik}\delta_{jl} + \delta_{il}\delta_{jk}\big)\delta(t - t'). \qquad (4)$$

The weight functions $\omega^D$ and $\omega^R$ must be related as

$$\omega^D(r_{ij}) = [\omega^R(r_{ij})]^2, \tag{5}$$

while the constants in equations (1) and (2) must obey the equality

$$k_B T = \frac{\sigma^2}{2\gamma} \tag{6}$$

for the equilibrium probability distribution function of DPD to be the canonical distribution function [4]; $k_B$ is the Boltzmann's constant and $T$ the absolute temperature. Although there is in principle no restriction on the spatial dependence of the functions in equations (1) and (2), as long as equations (5) and (6) are obeyed, it is customary to choose them *only* for computational convenience as equal to:

$$\omega^R(r_{ij}) = max\left\{\left(1 - \frac{r_{ij}}{r_c}\right), 0\right\}. \tag{7}$$

In addition to the viscous and random forces presented in equations (1) and (2), respectively, there is a conservative force in the DPD model, given by the following equation:

$$\boldsymbol{F}^C_{ij} = a_{ij}\left(1 - \frac{r_{ij}}{r_c}\right)\hat{\boldsymbol{r}}_{ij}, \tag{8}$$

with all three forces (equations (1), (2), (8)) becoming zero when $r_{ij} > r_c$. The factor $a_{ij}$ in equation (8) represents the strength of the conservative force between particles *i* and *j*, and accounts for all the thermodynamic properties of a pure DPD fluid [14]. Equation (8) represents the local excluded volume interactions of the fluid, see Fig. 1, and although it allows for particle – particle overlap, this does not occur in practice because the factor $a_{ij}$ in equation (8) is proportional to $k_B T/r_c$, which is very large (~ 78 for coarse – graining degree equal to 3, in reduced DPD units) and the overlapping of particles is therefore thermodynamically unfavorable [15].

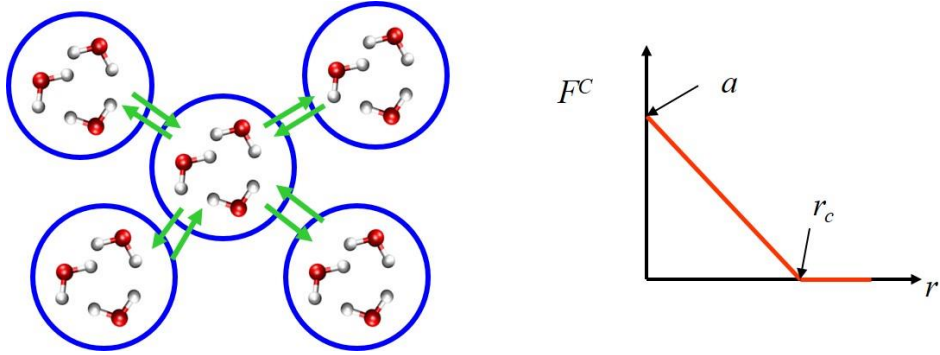

**Figure 1**. Schematic representation of the magnitude of the conservative DPD force ($F^C$). The image on the left represents a coarse – graining degree equal to three water molecules per DPD bead, while the one on the right is the spatial dependence of such force for a simple homogeneous fluid, see equation (8).

The physical meaning of the viscous and Brownian forces is similar to that in other types of dynamics with dissipative forces such as Langevin [16] or Brownian dynamics [17], with the important exception that, by definition, equations (1) and (2) obey explicitly Newton's third law of motion, whereas that is not the case with other comparable models [6]. Figure 2 illustrates schematically the competition and cooperation between the forces given by equations (1) and (2).

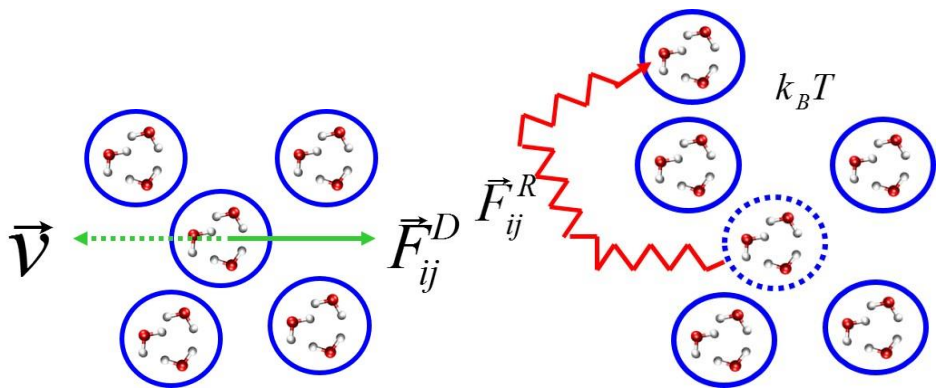

**Figure 2**. The left image represents the viscous force ($F^D$) acting in the collision of two DPD particles with relative velocity $v$ while the right image is a schematic representation of the Brownian motion ($F^R$) between any two particles, at a given temperature $T$. The circles are the DPD beads, which in the present work represent the volume of three water molecules. Provided here only for illustrative purposes. See text for details.

If the forces defined in equations (1) and (2) are coupled so that equations (5) and (6) are fulfilled, then the DPD system is assured to evolve to an equilibrium state with probability distribution defined by a fixed particle number ($N$), at the given volume ($V$) and temperature ($T$), i.e. by the canonical distribution function [4]. Additionally, all three forces (equations (1), (2) and (8)), have short range, which allows one to use large finite time steps when solving the equation of motion of the particles in numerical simulations. This fact, coupled with the coarse – graining procedure involved in the calculation of the conservative force parameter between any pair of particles $i$ and $j$, $a_{ij}$ (see equation (8)), make of DPD a mesoscopic scale model. There is a well – established algorithm to obtain these conservative force parameters [14] for molecules with specific chemical structure. Also, their dependence with temperature has been developed and tested successfully for the prediction of the interfacial tension and scaling properties of binary mixtures of immiscible fluids [13], but those topics are beyond the scope of the present work.

**THE VIRIAL THEOREM IN THE DPD MODEL**

One of the most frequently used routes to calculate the equation of state of complex fluids with computer simulations is the so – called virial theorem method [2]. The pressure is obtained from

$$P = \rho k_B T + \frac{1}{3V} \langle \sum_{j>i} \vec{F}_{ij} \cdot \vec{r}_{ij} \rangle, \qquad (9)$$

where $\rho$ is the number density, $\vec{F}_{ij}$ is the sum of forces (1), (2) and (8), and the brackets represent the time average. The first term in the right – hand side of equation (9) is the kinetic energy contribution to the total pressure, which is just the ideal gas contribution. The second term in the right – hand side of equation (9) is the so – called "excess pressure" [2], which contains all the effects of the non – ideal interactions. Essentially all works on DPD use only the conservative force when calculating the pressure of the fluid through equation (9), assuming that the dissipative and random forces do not contribute. When that is the case Groot and Warren [14] have shown that a pure DPD fluid obeys the following equation of state:

$$P = \rho k_B T + \alpha a \rho^2 , \qquad (10)$$

which is valid for $\rho > 2$, $\alpha$ is a constant found numerically to be $\alpha = 0.101 \pm 0.001$, and $a$ is the value of the conservative force constant, see equation (8) [14]. However, when the thermostat (dissipative and random) forces were included in the calculation of the pressure via equation (9), Groot and Warren [14] found differences of about 0.7 % with respect to the case when only the conservative force was considered in (9), without providing reasons for this admittedly small but measurable difference. Here we provide mathematical and computational arguments to determine when one can use only conservative forces when using equation (9) for the calculation of the pressure.

Let us start by considering the contribution of the random force, equation (2), to the pressure in equation (9). Using the fact that the weight function $\omega^R(r_{ij})$, equation (7), has an upper bound equal to 1 and is never negative, and that the maximum correlation length between any two given particles is $r_{ij}^{max} = r_c$ one can write the virial contribution of the random force as follows

$$0 \leq \langle \sum_{j>i} \vec{F}_{ij}^R \cdot \vec{r}_{ij} \rangle \leq \frac{\sigma r_c}{4 t_{Tot}} \sum_{j>i} \int_0^{t_{Tot}} dt \xi_{ij}(t) \qquad (11)$$

where $t_{Tot}$ is the total simulation time. For a sufficiently large value of $t_{Tot}$, and using the properties of the random numbers $\xi_{ij}$ given by equations (3) and (4) the integral in the inequality (11) can be made as small as needed, and we arrive at conclusion that the Brownian force does not contribute to the total pressure of the system, given by equation (9). For the evaluation of the contribution of the dissipative force, equation (1), to the pressure it is important to emphasize that this force depends on the relative velocity between the pair of particles colliding. Then, its contribution to the pressure can be written as

$$\langle \sum_{j>i} \vec{F}_{ij}^D \cdot \vec{r}_{ij} \rangle = -\frac{\gamma}{t_{Tot}} \sum_{j>i} \int_{r_{ij}(0)}^{r_{ij}(t_{Tot})} dr_{ij}\, \omega^D(r_{ij}) r_{ij}, \qquad (12)$$

which also tends to zero as the simulation time $t_{Tot}$ is increased because the integral in equation (12) is bound since the system of particles is a condensed fluid phase; recall from equations (5) and (7) that $\omega^D(r_{ij}) = (1 - r_{ij}/r_C)^2$, for $r_{ij} \leq r_C$. However, it is crucial to have a dissipative force that not only depends on the relative velocity between colliding particles, but that transfers the energy dissipated in such collisions to Brownian motion so that the kinetic energy is not lost, otherwise the system would eventually freeze. Hence the conditions given by equations (5) and (6) are indispensable for the negligible contribution to the total pressure of the dissipative and random forces in DPD for sufficiently long simulations.

To test (11) and (12) we have performed canonical – ensemble DPD simulations of a simple fluid made up of monomeric, identical particles using a conservative force constant $a_{ij} = 78.3$, see equation (8), with dissipative and random force constants equal to $\gamma = 4.5$ and $\sigma = 3$, respectively (see equations (1) and (2)) so that $k_B T = 1$, as stated in equation (6), unless stated otherwise. The equation of motion is solved numerically using an adapted form of the velocity – Verlet algorithm [18], for different choices of time step, in cubic boxes of various lengths and particle number so that the density could be fixed at three different values, $\rho =$ 3, 4 and 5. Periodic boundary conditions were imposed on all faces of the simulation box. We use reduced DPD units throughout this work, unless stated otherwise.

In Fig. 3 we show the evolution with time of the three DPD virial contributions to the pressure of a liquid made up of 3000 particles in a cubic box with side length equal $L^* = 10$ and periodic boundary conditions, for three different choices of the time step used in the integration of the equation of motion. Fig. 3(a) corresponds to the random force virial and Fig. 3(b) is the one corresponding to the dissipative force; both are shown to decrease to zero as the time of the simulation is increased, as fully expected from (11) and (12), respectively. However, while the contribution of the dissipative force to the virial decays monotonically as the number of time steps is increased, the random force contribution is somewhat noisy because of its very definition, see equation (2), but it quickly becomes negligible. The dissipative force contribution to the virial takes somewhat longer to decay, especially for small time steps. This feature may lead to artifacts in the calculation of the pressure if the total simulation time is not sufficiently long, or if the time step is not big enough. For comparison we have included in Fig. 3(c) the contribution of the conservative force to the virial for the same system, where we observe that a constant value is reached at long simulation time. At the smallest integration time step ($\Delta t = 0.0001$), the conservative force contribution to the virial takes longer to reach its equilibrium value, as expected.

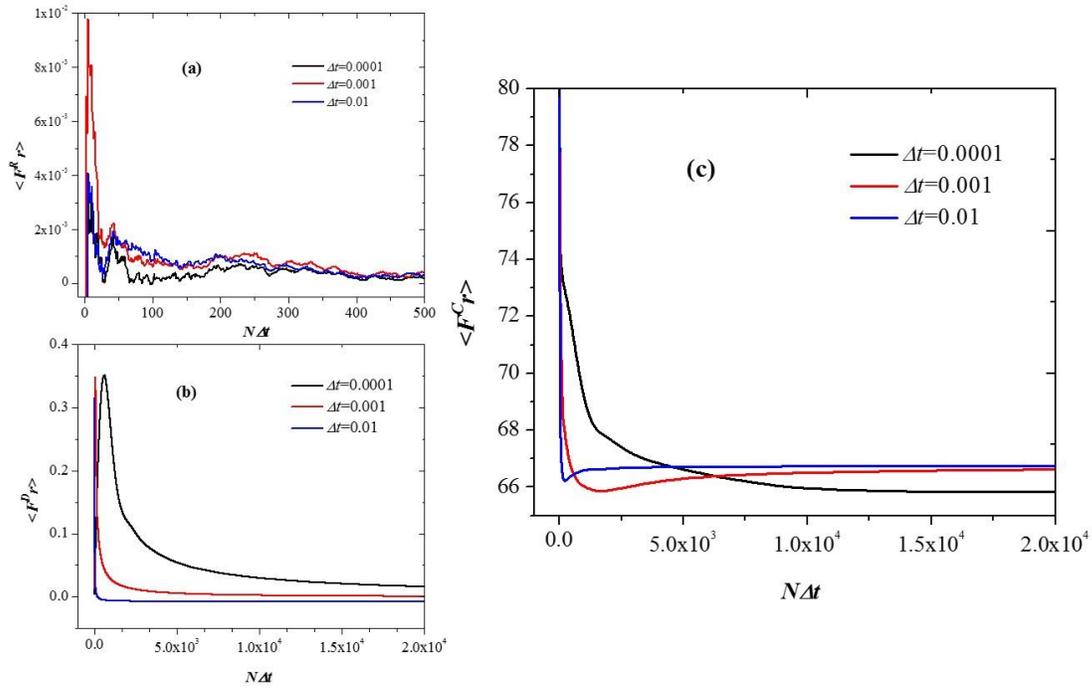

**Fig. 3 (a)** The virial contribution of the random and **(b)** dissipative forces to the pressure, see equation (9) for a DPD simple fluid as a function of time, for three choices of the integration time step. **(c)** The virial contribution of the conservative force with constant $a_{ij} = 78.3$. The fluid is made up of 3000 identical particles in a cubic box with volume $V=10\times10\times10$. All axes are shown in reduced DPD units.

We have carried out simulations with increasing values of the volume of the box while keeping the density constant ($\rho^*=3.0$) to check that these general trends do not depend strongly on finite size effects. In Fig. 4 we show the virial contribution of the three different types of DPD forces to the virial for four different volumes of the cubic simulation box, fixing the integration time step at the largest value shown in Fig. 3, namely $\Delta t=0.01$. All average virial contributions converge quickly to their equilibrium values as the box volume increases. Comparison of Fig. 4 with Fig. 3 shows that the influence of finite size effects is much weaker than the choice of time step on the average contribution of the DPD forces to the virial theorem, which is not surprising since the forces are short ranged. Figs. 4(a) and 4(b) show also that the average virial contribution to the pressure from the random and dissipative forces, respectively, is essentially zero for long simulations, as shown in Figs. 3(a) and 3(b) when $\Delta t =0.01$.

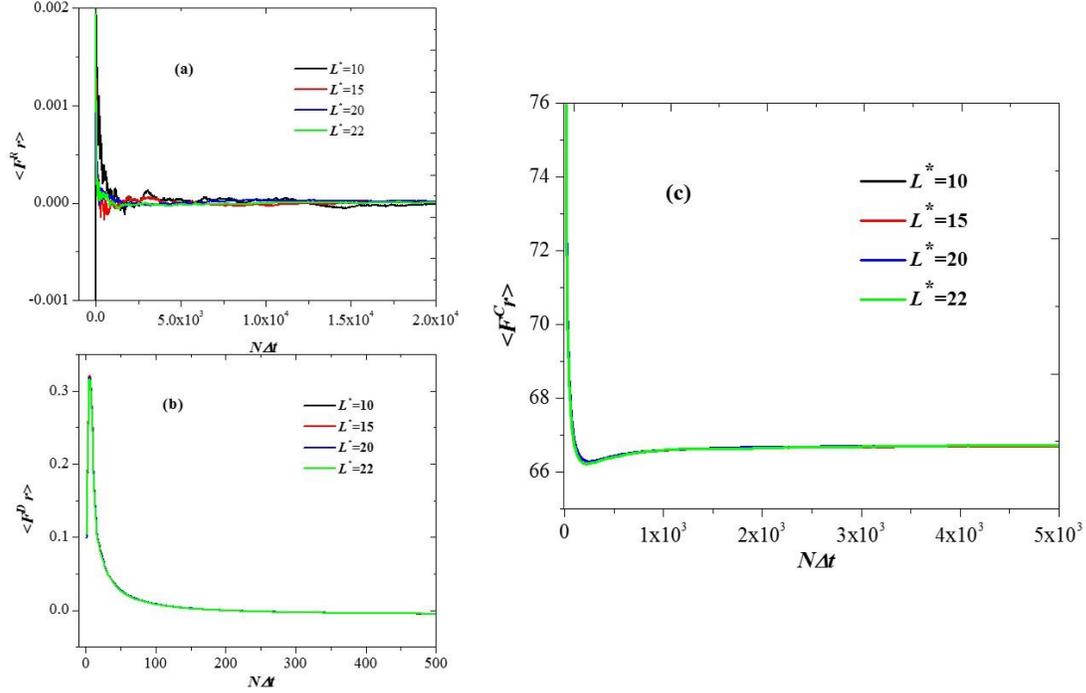

**Fig. 4** Effect of the size of the simulation box in the virial contribution of the **(a)** random, **(b)** dissipative and **(c)** conservative forces to the pressure, see equation (9) for a DPD fluid as a function of time, for four values of the volume of the cubic box with side length $L^*$. The conservative force constant is $a_{ij} = 78.3$. The fluid density is kept equal to $\rho^* = 3$ and the integration time step was chose as $\Delta t=0.01$ in all cases. All quantities are shown in reduced DPD units.

The balance between the dissipative force and the random force make of the DPD thermostat one of the most stable ones, particularly for applications to non – equilibrium simulations [19]. Most DPD works *assume* that the strength of the frictional and Brownian forces is small enough so that they have negligible influence on the dynamics of the system, but this has not been explored. To determine the extent to which the strength of the frictional and random forces introduces artifacts into the equilibrium value of the pressure we have carried out simulations with various values of $\sigma$ (amplitude of the stochastic force) and $\gamma$ (strength of the viscous force), while keeping the relation between them fixed, as dictated by the fluctuation – dissipation theorem, i. e., $\sigma^2/2\gamma = k_B T$. The contribution of the three forces to the virial for these cases at the same density ($\rho^*=3$) can be found in Fig. 5, and because equation (6) is always fulfilled the temperature in all cases is equal to one, in reduced units. The case with $\sigma=1$ and $\gamma=0.5$ corresponds to a fluid with very little viscosity and Brownian motion, while that with $\sigma=4$ and $\gamma=8$ models a very viscous fluid with many Brownian collisions between the particles. In Fig. 5(a) one notices that the least viscous fluid takes longer to reach a vanishing contribution of the random force to the virial, because the thermostat is too weak to act quickly and requires of many collisions between the particles. In such case, if one were to run simulations in blocks of, say, $10^4$ time steps as is usually done, there would be a small but finite artificial contribution to the total pressure arising from the random force. For the more viscous types of fluids seen in Fig. 5(a) this is not a problem and the system quickly arrives at the equilibrium. In Fig. 5(b) we show the contribution of the dissipative force to the virial for the same cases seen in Fig. 5(a); although in all cases the contribution of the dissipative force to the virial decays quickly to zero, the case with $\sigma=1$ and $\gamma=0.5$ is the one with the smallest contribution, precisely because $\gamma$ is so small. However, the largest influence of the various values of the strength of the random and dissipative forces is found in the conservative force contribution to the virial, which is the most important one since it is the one used for the calculation of the pressure of the fluid, and it is shown in Fig. 5(c). The contribution of what may be called the weakest thermostat ($\sigma=1$, $\gamma=0.5$) takes the longest to arrive at the equilibrium state; in fact, even after $10^5$ time steps it has not yet done so. Increasing the "strength" of the thermostat improves the ability of the fluid to reach equilibrium, and in those cases the calculation of the pressure would be free from these artifacts, for sufficiently long simulations. The results shown in Fig. 5 underscore the importance of the careful assessment of the role of viscous and stochastic forces in numerical simulations.

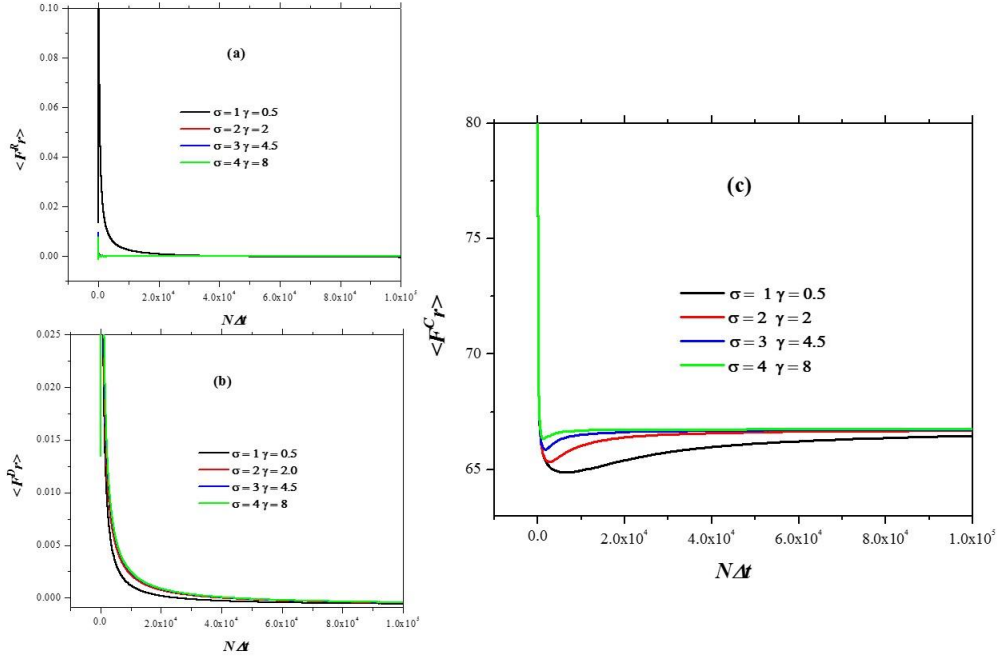

**Fig. 5** Effect of varying the strength of the random ($\sigma$) and dissipative ($\gamma$) forces in the virial contribution of the **(a)** random, **(b)** dissipative and **(c)** conservative forces to the pressure, see equation (9) for a DPD fluid as a function of time. In all cases equation (6) is obeyed, thus the temperature is always $T^*=1$. The conservative force constant is $a_{ij} = 78.3$. The fluid density is kept equal to $\rho^* = 3$ and the integration time step was chose as $\Delta t=0.001$ in all cases. All quantities are shown in reduced DPD units.

To further illustrate the importance of the fluctuation − dissipation theorem in DPD we performed simulations where the random force was set equal to zero at the beginning of the simulation and the system was allowed to evolve to a state where all the particles' positions where eventually frozen, yielding an exceedingly small value of the kinetic energy, and therefore of the temperature.

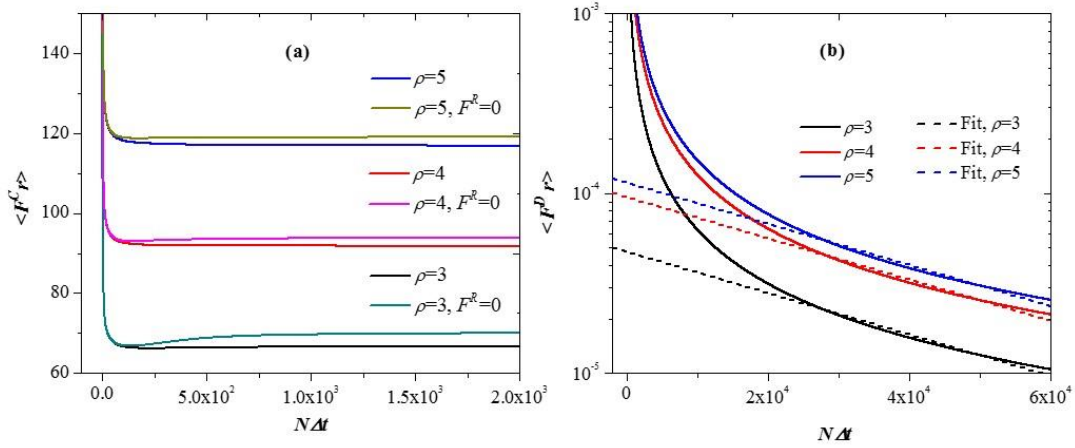

**Fig. 6 (a)** Comparison between the average virial contribution of the conservative force for a simple DPD fluid obtained with and without the random force in the system, the latter was done by setting $\sigma=0$, for three different values of the density. **(b)** Average contribution of the dissipative force to the virial when there is no random force, for the same densities as in (a). The dotted lines are the best fits for the long time tail of the curves to the function $\exp(-t/\tau)$; see text for details. The integration time step was set to $\Delta t=0.01$ in all cases. Axes are shown in reduced DPD units.

Fig. 6(a) displays the average value of the virial contribution of the conservative force for systems of particles at various densities where the random force was set initially to zero, compared with the equivalent cases where the random force obeys the fluctuation – dissipation theorem, see equations (5) and (6). As the figure shows, a constant value of the virial contribution is achieved, with the value of the frozen system being larger than the one with non - zero kinetic energy. This is to be expected since the virial theorem establishes that:

$$\langle \sum \overrightarrow{F_{ij}^C} \cdot \overrightarrow{r_{ij}} \rangle = 3VP - 3V\langle KE \rangle, \qquad (13)$$

where $V$ is the volume of the simulation box and $\langle KE \rangle$ is the average kinetic energy. Hence, for fluids with particles in motion and $\langle KE \rangle > 0$, the right – hand side of equation (13) should yield a lower value of the virial as that compared with frozen systems, where $\langle KE \rangle = 0$, in agreement with equation (13) and Fig. 6(a). As the density is increased once notices the difference between the average virial contribution of the frozen and moving systems becomes smaller, which is due to the reduced degrees of freedom of the particles when the density is increased, which in turn reduces the space available for particle displacements and the hence the kinetic energy of the particles. Fig. 6(b) shows the behavior of the average virial contribution of the dissipative force when the frictional force is set to zero for the same densities as in Fig. 6(a). The inertial decay of the motion of the particles that make up the fluid is shown to be exponentially decreasing. The fluid's viscous force is gradually smaller because the relative velocity of particles colliding is also decreasing since there is no Brownian force to sustain the dissipation. The processes shown in Fig. 6(b) are clearly not any more equilibrium states, rather there is "slowing down" of the particles of the fluid whose correlations are decaying as $e^{-t/\tau}$ with a characteristic decay time given by $\tau = \gamma r_C/a$, where $a$ is the conservative force constant (see equation (8)), $r_C$ is the characteristic DPD length, and $\gamma$ is the effective strength of the dissipative force. The latter cannot be its equilibrium value $\gamma = 4.5$, see equation (1), because if there is no frictional force the fluctuation – dissipation theorem ceases to operate in the fluid. Instead, $\gamma$ is the amplitude of the effective viscosity that the DPD particles experience when they lose their kinetic energy. Therefore the relation $\gamma r_C/a$ can be interpreted as the competition between the viscous force and the conservative force as the fluid is frozen.

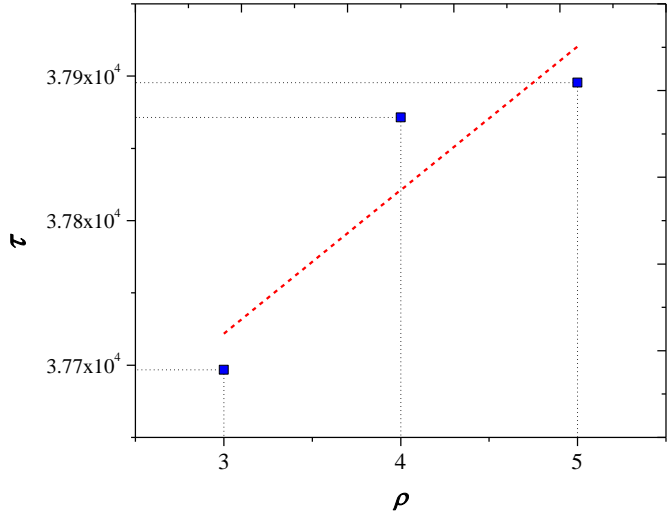

**Fig. 7** The characteristic decay time of the average contribution to the virial of the dissipative force, obtained from the best fits in Fig. 6(b), as a function of the density. The dotted line is a linear fit. The integration time step was $\Delta t=0.01$. The axes are shown in reduced DPD units.

The characteristic time extracted from the fits shown in Fig. 6(b), which we have identified as $\tau = \gamma r_c/a$, is shown in Fig. 7 as a function of the density of the DPD fluid and is rather large because it corresponds to a large effective viscosity $\gamma$ that appears as a result of having removed the random force. Alternatively, the fluid can be thought of as having a very small diffusion – like coefficient, $D$, behaving as $D \sim 1/\gamma \sim 1/\tau$, and according to the trend seen in Fig. 7, $D \sim 1/\rho$, which is in agreement with the diffusion coefficient predicted earlier [14, 20] for a simple DPD fluid with non – zero random force, $D = 45 k_B T / 2\pi\gamma\rho r_C^3$. The same general conclusions we have reached here can be obtained if one performs the average in equation (9) over an ensemble rather than over time, as it is the case when one carries out Monte Carlo (MC) simulations [21].

**CONCLUSIONS**

Coarse – grained simulations that incorporate dissipative forces and Brownian motion like DPD have proved to be very successful tools in predicting the behavior of physicochemical phenomena in complex fluids. A common route to accomplish a thorough understanding in these cases is the detailed knowledge of the equation of state of the model, which requires usually the application of the well – known virial theorem for the calculation of the total pressure of the system, arising from the forces acting in the system. Here we have shown that the dissipation and random forces that make up the DPD model do not contribute to the pressure of the system and thereby they can be neglected in the prediction of the thermodynamic behavior of complex fluids. However, for this statement to be true the viscous force must be proportional to the relative velocity between colliding particles and, most importantly, such force must be balanced with the Brownian force according to the fluctuation – dissipation theorem. If the latter condition is not met, which occurs for example when there is no random force, the fluid will evolved to a frozen state where the dissipative force consumes all the available kinetic energy in the system and the particles' motion decays with an effectively decreasing diffusion – like coefficient. Based on these results one can conclude that using equation (9) with conservative forces only in equilibrium DPD simulations in the canonical ensemble should lead to the correct thermodynamic state as long as the fluctuation – dissipation theorem is obeyed by the dissipative and random forces.

## ACKNOWLEDGEMENTS

AGG and MABA would like to thank Universidad Autónoma de San Luis Potosí for the hospitality and necessary support for this project. JDH thanks also the Polymer Group at IFUASLP. The authors acknowledge R. Catarino Centeno for discussions.